\documentclass[journal]{IEEEtran}

\IEEEoverridecommandlockouts 

\usepackage{subfigure}
\usepackage{amsmath}
\usepackage{graphicx}
\usepackage{subfigure}
\usepackage{amssymb}
\usepackage{array}
\usepackage{mathtools}
\usepackage{bbm}

\title{\LARGE \bf A Two Level Feedback System Design to Regulation Service Provision}

\author{Bowen Zhang$^{1}$ and John Baillieul$^{2}$
\thanks{*The authors gratefully acknowledge support of the U.S. National Science Foundation under EFRI Grant 1038230}
\thanks{$^{1}$Corresponding author. Division of Systems Eng., Boston University, 15 St. Mary’s St., Brookline, MA 02446, email:{\tt\small bowenz@bu.edu}}%
\thanks{$^{2}$Dept. of Electrical and Computer Eng., Dept. of Mechanical Eng., and Division of Systems Eng., Boston University, 110 Cummington St., Boston, MA 02215, email:{\tt\small johnb@bu.edu}}%
}

\begin{document}
\maketitle
\thispagestyle{empty}
\pagestyle{empty}
\begin{abstract}
Demand side management has gained increasing importance as the penetration of renewable energy grows. Based on a Markov jump process modelling of a group of thermostatic loads, this paper proposes a two level feedback system design between the independent system operator (ISO) and the regulation service provider such that two objectives are achieved: (1) the ISO can optimally dispatch regulation signals to multiple providers in real time in order to reduce the requirement for expensive spinning reserves, and (2) each regulation provider can control its thermostatic loads to respond the ISO signal. It is also shown that the amount of regulation service that can be provided is implicitly restricted by a few fundamental parameters of the provider itself, such as the allowable set point choice and its thermal constant. An interesting finding is that the regulation provider's ability to provide a large amount of long term accumulated regulation and short term signal tracking restrict each other. Simulation results are presented to verify and illustrate the performance of the proposed framework. 
\end{abstract}

\section{Introduction}
\label{introduction}

Wind energy, known as the most widely used form of clean and renewable generation resource in power systems, is increasing its penetration around the world. Europe, Asia, and North America will have an exponential yearly increase in the amount of added wind power capacity by 2020 \cite{emerge}. The increase in wind energy makes the generation side less controllable due to the intermittent feature of wind. As a consequence, building level demand side management (DSM) for thermostatic load becomes a crucial part that can be achieved generally through either a price-based signalling control \cite{caram1}-\cite{albadi} or a non-priced \textit{direct load control} (DLC). Compared with pricing mechanism, the operator in DLC can have better understanding of the loads' response in order to provide short time scale regulation services \cite{duncan1}.

DLC approaches on thermostatic loads have been studied extensively during the past decade for difference purposes, including load shifting, load smoothing, and load following. A primitive investigation on load shifting has been studied with a state queuing model to illustrate the effect of set-point change \cite{chassin1}. It is shown that even a simple step change will result in a complicated load profile due to the instantly change of loads' status. Taking into account both the appliances duty cycle as well as probabilistic consumer behaviour, a steady state appliances population distribution during an load shifting demand response is analysed in \cite{chassin2}. Results shows that the steady state distribution exists and may not be as simple as uniform. More recently, the possibility to integrate DLC into load smoothing or following is investigated. The notion of packetized DLC is proposed in \cite{bowen} to smooth electricity consumption with shortened appliances duty cycle. Load following can also be found in \cite{duncan2}-\cite{duncan4} where various of control methods including discrete time directly on/off actuation control, set point shifting, as well as model based predictive control are used. Results in these papers show that loads, by acting as both a positive and negative generation source, is a promising resource to response against the load request from the ISO to maintain supply demand balance. 
 
While the study of the DLC itself seems to be comprehensive, there are two issues remain to be answered. The first one is the limitation of regulation provision, namely what is the maximum regulation can be provided by a specific provider given its implicit features such as population of appliances pool, parameters in its own model. The second one is about the information exchange between the ISO and the providers. Instead of the one way command-like signalling, can we design another framework such that the ISO can get feedback information from providers? If so, how beneficial is it to the real time operation?

The objective of this paper is to answer these questions by developing a two layer feedback control system. We focus primarily on smart buildings that are operated by a central operator who has the authority to shift the set point of thermostatic appliances within the building. Consumers are assumed to authorize the operator to control their set point once they provide the allowable set point range. It will be shown that the operator in the lower feedback layer can design a feedback control law to track the ISO regulation signal within a certain ramping rate. The higher feedback loop is designed between the ISO and buildings where the former dispatches real time signals and the latter send back information to characterize their instant-by-instant capability to respond. We also derive analytically the maximum possible regulation that can be provided from buildings based on two considerations -- the capability to provide long term accumulated and short term (one step) ramping regulation. It is further shown that two fundamental parameters determining the capability are the building thermal constant $\tau$ and appliances temperature gain $T_{\textrm{g}}$. Large $\tau$ enables the building to provide larger long term, but smaller short term regulation service. Large $T_{\textrm{g}}$ performs oppositely. The overall two level system has a good performance in real time operation as it reduces the amount of spinning reserves required from high-cost generators.

The paper proceeds as follows. Section \ref{markov jump process model} develops the state space model for which the feedback system is designed. Section \ref{two level feedback system design} gives the overall design of the two layer feedback system in which we first introduce the lower layer controller with feedback linearization, followed by the investigation of maximum possible regulation provision, and then the higher level feedback design that includes a quadratic program to optimally dispatch real time regulation to individual buildings. Section \ref{state observer design} discusses an observer design to asymptotically reduce the estimation error for unobservable states. Simulation results are given in section \ref{simulation section} to show the performance of the proposed system. Section \ref{conclusion section} concludes the paper and proposes future work.

\section{Markov Jump Thermal Process Modelling}
\label{markov jump process model}
We model the system as a continuous time, discrete state Markov jump process where a state $i$ is defined as a pair of temperature value and thermostat value $\{T_{i},\textrm{on(off)}\}$. Previous literature has considered similar state definitions - for example \cite{chassin1}, \cite{chassin2}, \cite{duncan2}. The departure of our modelling approach from previous ones is that we consider state in a Markov setting and establish the implicit relations between thermal processes and the Markov jump processes. The Markov jump process model in this section is the fundamental model that we based to design the two level feedback system and to derive the limitation of providing regulation service.

Denote the fixed width comfort band around the set point as $[T_{\textrm{min}},T_{\textrm{max}}]$ and discretize temperature in the band into bins of width $\delta$, then the number of temperature bins is $N=\Delta_{\textrm{band}}/\delta$. We say that an appliances is in state $i$ for $i=1,\ldots,N$ if $T_{i}\in [T_{\textrm{min}}+(i-1)\delta,T_{\textrm{min}}+i\delta]$ with status off, and $i$ for $i=N+1,\ldots,2N$ if $T_{i}\in [T_{\textrm{max}}-(i-N-2)\delta,T_{\textrm{max}}-(i-N-1)\delta]$ with status on.

First let our control $u=0$ and solve for the Markov transition rate of the uncontrolled thermal process. Denote $\alpha$ the transition rate when the thermostat is off, and $\beta$ the rate when the thermostat is on; see Fig.1$(\textrm{a})$. In the duty off process, the probability that an appliances will be in state $i$ at time $t+1$ given in state $j$ at time $t$ is given by,
\begin{equation}
\label{def_rate}
p_{i,j}= \left\{
\begin{array}{ll}
\alpha \Delta t + o(\Delta t) & \textrm{if $i=j+1$} \\
1-\alpha \Delta t + o(\Delta t) & \textrm{if $i=j$} \\
0 & \textrm{otherwise},
\end{array}
\right.
\end{equation}
where $o(\Delta t)$ means $\lim\limits_{\Delta t\rightarrow 0} \frac{o(\Delta t)}{\Delta t}=0$. The above equation relates transition probability and transition rate for small $\Delta t$. It can be interpreted that the transition probability between adjacent states linearly increases with small $\Delta t$. 

\begin{figure}[htb]
\centering
\label{trans1}
\includegraphics[width=0.4\textwidth,height=0.2\textheight]{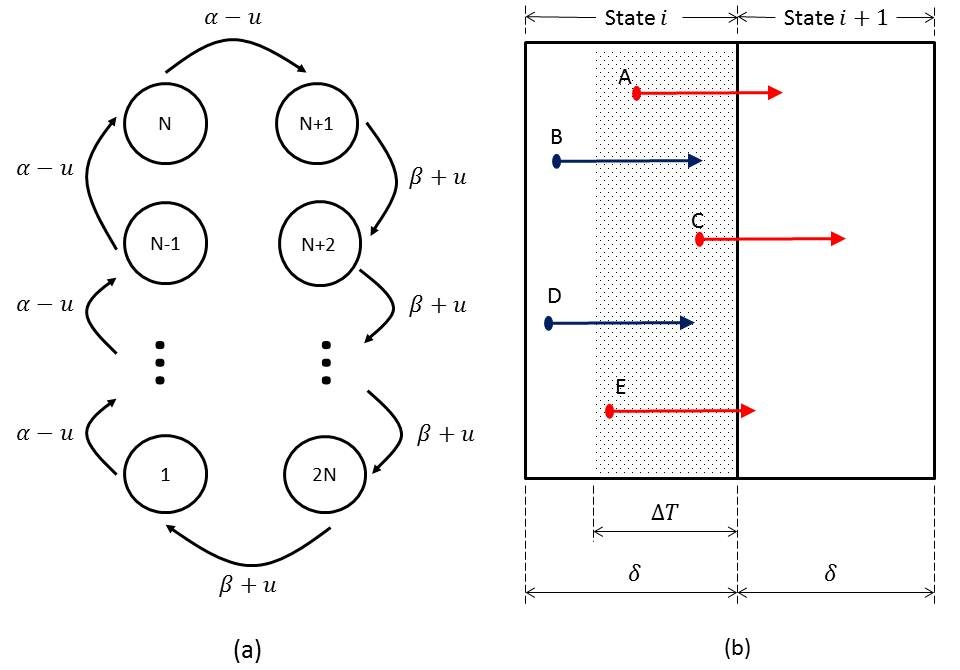}
\caption{Markov jump process modelling. (a) Markov chain transition rate diagram. (b) Transition from state $i$ to $i+1$. With temperature rise $\Delta T$ (length of arrow), appliances whose temperatures inside the dotted area (A, C, E) change state and outside (B, D) remain in the same state.}
\end{figure}

From a thermal point of view, the temperature rise $\Delta T$ within a small $\Delta t$ is proportional to its warming rate $r_{\textrm{off}}$,
\begin{equation}
\Delta T = r_{\textrm{off}}\Delta t.
\end{equation}
Assuming that an appliance's actual temperature is uniformly distributed among its bins, then the probability that the state transits is equal to the probability that the appliance temperature is in the dotted area in Fig.1$(\textrm{b})$, namely appliances in the dotted area change their state from $i$ to $i+1$, and appliances outside remain in the same state.

The probability of being in the dotted area is given by,
\begin{equation}
\label{probtrans}
p_{\textrm{dot}}=\frac{r_{\textrm{off}} \Delta t}{\delta}=\frac{\Delta_{\textrm{band}}\Delta t}{t_{\textrm{off}}\delta}=\frac{N\Delta t}{t_{\textrm{off}}},
\end{equation}
where the second equality follows from the relation between warming rate and duty cycle. Since $p_{i,i+1}=p_{\textrm{dot}}$, comparing (\ref{def_rate}) with (\ref{probtrans}) we have $\alpha=N/t_{\textrm{off}}$. Similarly, $\beta=N/t_{\textrm{on}}$ for the duty off process. 

When control is applied to shift the set point at a rate of $r_{\textrm{set}}$ (the unit of $r_{\textrm{set}}$ is the same as $r_{\textrm{on}}$ or $r_{\textrm{off}}$), there is also a transition between adjacent states within $\Delta t$. Intuitively, when we change the set point, the absolute temperature in each individual room does not change instantly, but its relative position in the comfort band changes. When the set point rises, namely $r_{\textrm{set}}>0$, we can show that the resulting transition rate is $u=r_{\textrm{set}}/\delta$. The transition rate is the same with $r_{\textrm{set}}<0$. The combined transition rate by the thermal process and set point shifting process is the sum of these two individual processes as shown in Fig.1$(\textrm{a})$. Note that the allowable control set is $\textit{L}_{\textrm{u}}=\{u|-\beta \leq u \leq \alpha\}$ to maintain a non-negative Markov rate. When $u>\alpha$ or $u<-\beta$, the system fails to be a Markov chain. 

Similar to (\ref{def_rate}), we can write the transition probability for the duty off process with non-zero control $u$,
\begin{equation}
\label{trans_prob}
p_{i,j}= \left\{
\begin{array}{ll}
(\alpha-u) \Delta t + o(\Delta t) & \textrm{if $i=j+1$} \\
1-(\alpha-u) \Delta t + o(\Delta t) & \textrm{if $i=j$} \\
0 & \textrm{otherwise}.
\end{array}
\right.
\end{equation}
Let $x(t)$ be a vector whose $i^{th}$ component is the probability that an appliance is found in the $i^{th}$ state in the Markov chain. The dynamics of $x_{\textrm{i}}(t)$ for $i=2,\ldots,N$ is given by,
\begin{equation}
\label{555}
x_{i}(t+\Delta t)=p_{i,i}x_{i}(t)+p_{i,i-1}x_{i-1}(t),
\end{equation}
Substitute (\ref{trans_prob}) into (\ref{555}) we will have,
\begin{equation}
x_{i}(t+\Delta t)-x_{i}(t)=(\alpha-u)\Delta t [x_{i-1}(t)-x_{i}(t)] +o(\Delta t).
\end{equation}
Dividing $\Delta t$ on both sides and taking $\Delta t$ to 0 yields the dynamics for the $i^{th}$ component in the duty off process,
\begin{equation}
\label{1st component}
\dot{x}_{i}(t)=(\alpha-u)[ x_{i-1}(t)-x_{i}(t)], 
\end{equation}
for $i=2,\ldots,N$. Similarly for the duty on process $i=N+2,\ldots,2N$,
\begin{equation}
\label{2nd component}
\dot{x}_{i}(t)=(\beta+u) [x_{i-1}(t)-x_{i}(t)], 
\end{equation}
and reflection boundaries $i=1,N+1$,
\begin{equation}
\label{3rd component}
\begin{array}{ll}
\dot{x}_{1}(t)=-(\alpha+u) x_{1}(t)+(\beta+u) x_{\textrm{2N}}(t), \\
\dot{x}_{\textrm{N+1}}(t)=(\alpha-u) x_{\textrm{N}}(t)-(\beta+u) x_{\textrm{N+1}}(t). 
\end{array}
\end{equation}
From (\ref{1st component}) to (\ref{3rd component}), the overall dynamics of $x(t)$ is given by,
\begin{equation}
\label{main_dyn}
\dot{x}(t)=[A+Bu(t)]x(t),
\end{equation}
where
{\footnotesize
\begin{displaymath}
A = 
\left( \begin{array} {ccccccc}
-\alpha & 0 & &\ldots &  & & \beta \\
\alpha & \ddots & & & & & \\
0 & \ddots & -\alpha &\ddots &  & & \\
& & \alpha & -\beta &  & &\vdots \\
\vdots& & \ddots& \beta & \ddots & & \\
& & & &\ddots & -\beta & 0 \\
0 & &\ldots & & 0& \beta &-\beta
\end{array} \right),
\end{displaymath}
\begin{displaymath}
B = 
\left( \begin{array} {ccccccc}
1 & 0 & &\ldots &  & & 1 \\
-1 & \ddots & & & & & \\
0 & \ddots & 1 &\ddots &  & & \\
& & -1 & -1 &  & &\vdots \\
\vdots& & \ddots& 1 & \ddots & & \\
& & & &\ddots & -1 & 0 \\
0 & &\ldots & & 0& 1 &-1
\end{array} \right).
\end{displaymath}
}
The output of the system, namely the aggregated consumption, is
\begin{equation}
\label{main output}
y(t)=Cx(t),
\end{equation}
where $C=[\underbrace{0,\ldots,0}_{\textrm{N}},\underbrace{N_{\textrm{c}},\ldots,N_{\textrm{c}}}_{\textrm{N}}]$, and $N_{\textrm{c}}$ is the total number of appliances in the pool.

\section{Two Level Feedback System Design}
\label{two level feedback system design}
Based on the state space model developed in section \ref{markov jump process model}, we are able to design our two level feedback system as shown in Fig.2. In the building level feedback loop, we design a state feedback law such that the building can track the required regulation signal. The higher level feedback is from each individual building to the ISO where the former send their information to the latter to characterize the building's capability of responding to regulation signals. The ISO, after receiving all information from each participant, dispatches real time regulation signals after solving an optimization problem. The next sub-section will shortly discuss the building level controller design by feedback linearzation.
\begin{figure}[htb]
\centering
\label{Two_level_feedback}
\includegraphics[width=0.4\textwidth,height=0.2\textheight]{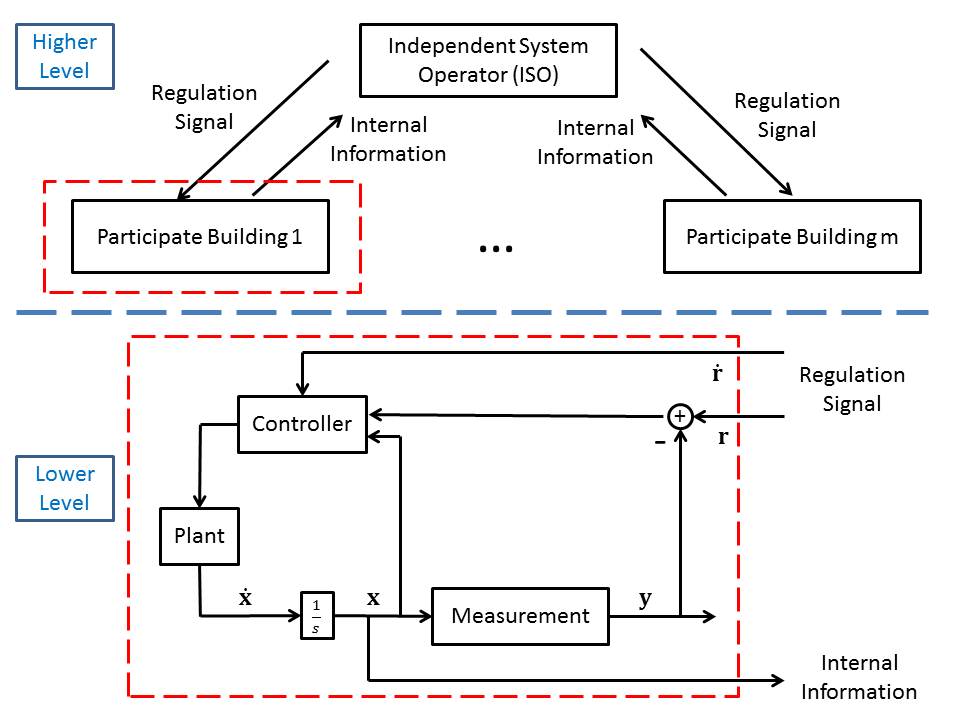}
\caption{Two level feedback system where the lower feedback designs the individual tracking controller for a given regulation signal, and the higher feedback enables the communication between the ISO and the regulation providers such that information is exchanged to reach better real time regulation signal dispatch.}
\end{figure}

\subsection{Building Level Feedback Linearization Design}
To solve the building level controller design for the non-linear system described in (\ref{main_dyn}) and (\ref{main output}), the feedback linearization method can be readily applied \cite{khalil}. Note that the system has relative degree \textbf{1} as the first derivative of $y$,
\begin{equation}
\dot{y}=C\dot{x}(t)= C[A+Bu(t)]x(t),
\end{equation}
depends on the control $u$ if  
\begin{displaymath}
\begin{array}{rl}
x(t) & \in \{x(t)\in\mathbb{R}^{\textrm{2N}}|CBx(t)\neq 0\},\\
& \in \{x(t)\in\mathbb{R}^{\textrm{2N}}|x_{\textrm{N}}(t)+x_{\textrm{2N}}(t) > 0\}.
\end{array}
\end{displaymath}
This means the number of controllable appliances around the comfort band boundary is positive, so shifting the set point can change the aggregated consumption. The following change of variables,
\begin{equation}
T(x)=\left[
\begin{array}{c}
\phi_{1}(x)\\
\vdots\\
\phi_{2n-1}(x)\\
Cx
\end{array}
\right] \coloneqq \left[
\begin{array}{c}
\eta\\
\xi
\end{array}
\right],
\end{equation}
yields the internal-external state space representation,
\begin{equation}
\begin{array}{lll}
\dot{\eta} & = & f_{0}(\eta(t),\xi(t))\\
\dot{\xi} & = & C[A+Bu(t)]x(t) \\
y & = & \xi(t) ,
\end{array}
\end{equation} 
where $\eta(t)$ stands for the internal and $\xi(t)$ the external states \cite{khalil}. Then the output becomes the external variable $\xi(t)$.

To design a controller for a relative degree 1 system such that $\lim\limits_{t\rightarrow\infty}\xi(t)-R(t)=0$ where $R(t)$ is the required amount of regulation needs to be provided, we need $R(t),\dot{R}(t)$ to be available and bounded for all $t>0$. This is obtainable since the ISO regulation requirement is bounded: $R(t)\in[-R_{\textrm{r}},R_{\textrm{r}}]$ ($R_{\textrm{r}}$ is maximum regulation sold to the market), and $R(t)$ is updated every 4 to 6 seconds prior to consumption. The time derivative of the tracking error,
\begin{equation}
\label{eeee}
\dot{e}(t)=\dot{\xi}(t)-\dot{R}(t)=C[A+Bu(t)]x(t) - \dot{R}(t),
\end{equation}
will asymptotically approach 0 if $u(t)$ is chosen such that
\begin{equation}
\label{stablee}
\dot{e}(t)=Ke(t), K>0.
\end{equation}
The controller
\begin{equation}
\label{controlaw}
u(t)=-\frac{CAx(t)}{CBx(t)}+\frac{1}{CBx(t)}[-K(\xi(t)-R(t))+\dot{R}(t)]
\end{equation}
satisfies the requirement because substituting (\ref{controlaw}) into (\ref{eeee}) yields (\ref{stablee}). For the discrete signal $R(t)$, we can calculate the first derivative of $R(t)$ as $\dot{R}(t)=\frac{\displaystyle R(t+\Delta t)-R(t)}{\displaystyle\Delta t}$.

\subsection{Long and Short Term Regulation Provision}
In this section we proceed to answer the first question proposed in section \ref{introduction}, namely what is the regulation service limitation that a building can provide given its model parameters and user specified settings. This question has two parts: $(1)$ in the long term, we need the accumulated shift of set point to be within the allowable range to provide reserve, and $(2)$ in the short term, we need the signalling change to be within bounds such that the controller can track it. The first proposition will discuss the limitation on long term regulation.

\textbf{Proposition.1} For a given number of appliances $N_{\textrm{c}}$ and allowable set point range $[T_{\textrm{set}}-\frac{1}{2}\Delta_{\textrm{set}},T_{\textrm{set}}+\frac{1}{2}\Delta_{\textrm{set}}]$, the accumulated amount of reserve that a building can provide up to time $t$ is given by
\begin{equation}
S(t)=\sum\limits_{i=0}^{t}R(i),
\end{equation}
and this is bounded by,
\begin{equation}
\label{thm1}
S(t)\leq \frac{N_{c}\tau\Delta_{\textrm{set}}}{2T_{\textrm{g}}},
\end{equation}
where $R(i)$ is the regulation signal at time $i$, $T_{\textrm{g}}$ is the temperature gain of air conditioner if it is on, and $\tau$ is the effective thermal constant of the building.

\textit{Proof.} The total consumption up to time $t$ is given by,
\begin{equation}
P_{\textrm{cool}}=tR_{\textrm{b}}+\sum\limits_{i=0}^{t}R(i)=tR_{\textrm{b}}+S(t),
\end{equation}
which is an average consumption level of 
\begin{equation}
P_{\textrm{ave}}=R_{\textrm{b}}+\frac{S(t)}{t}.
\end{equation}
This is equivalent to choosing a constant non-zero $u$ to maintain at a constant consumption level $P_{\textrm{ave}}$. The relation between average consumption and non-zero control is, 
\begin{equation}
P_{\textrm{ave}}=N_{\textrm{c}}\frac{t_{\textrm{on}}}{t_{\textrm{on}}+t_{\textrm{off}}}=N_{\textrm{c}}\frac{\frac{N}{\beta+u}}{\frac{N}{\alpha-u}+\frac{N}{\beta+u}}=N_{\textrm{c}}\frac{\alpha-u}{\alpha+\beta}.
\end{equation}
For an uncontrolled process at average consumption,
\begin{equation}
R_{\textrm{b}}=N_{\textrm{c}}\frac{\alpha}{\alpha+\beta}.
\end{equation}
From the above three equations, we have,
\begin{equation}
\frac{S(t)}{t}=N_{\textrm{c}}\frac{-u}{\alpha+\beta}
\end{equation}
resulting a constant value of control
\begin{equation}
\label{constctrl}
u=-\frac{S(t)(\alpha+\beta)}{tN_{\textrm{c}}}.
\end{equation}
The set point shift after time $t$ by the constant $u$ is given by,
\begin{equation}
\setlength{\extrarowheight}{7pt}
\begin{array}{ll}
T(t) & =T(0)+tr_{set}\\
&=T(0)+tu\delta \\
&=T(0)-\frac{\displaystyle S(t)}{\displaystyle N_{c}}(\frac{\displaystyle N}{\displaystyle t_{\textrm{off}}}+\frac{\displaystyle N}{\displaystyle t_{\textrm{on}}})\frac{\displaystyle T_{\textrm{band}}}{\displaystyle N}\\
&=T(0)-\frac{\displaystyle S(t)}{\displaystyle N_{c}}(r_{\textrm{on}}+r_{\textrm{off}})\\
&=T(0)-\frac{\displaystyle S(t)}{\displaystyle N_{c}}\frac{\displaystyle T_{\textrm{g}}}{\displaystyle\tau}.
\end{array}
\end{equation}
The third equation is based on the relation between transition rate and duty cycle. The last equation says that the sum of cooling and warming rate is the rate that the cooling system generates because $r_{\textrm{off}}=r_{\textrm{amb}}$ and $r_{\textrm{on}}=r_{\textrm{app}}-r_{\textrm{amb}}$, see \cite{bowen}. $r_{\textrm{amb}}$ is the warming rate caused by the ambient temperature that higher than room temperature, and $r_{\textrm{app}}$ is the cooling rate caused by the operation of air conditioner compressor. To maintain the set point shift within the allowable set point band with $T(0)=T_{\textrm{set}}$, we need
\begin{equation}
T(t)\in[T_{\textrm{set}}-\frac{1}{2}\Delta_{\textrm{set}},T_{\textrm{set}}+\frac{1}{2}\Delta_{\textrm{set}}],
\end{equation}
which yields
\begin{equation}
S(t)\leq \frac{N_{c}\tau\Delta_{\textrm{set}}}{2T_{\textrm{g}}}.
\end{equation}
$\blacksquare$

\textbf{Remark.1} According to proposition 1, the accumulated long term provision capability is proportional to three parameters: $N_{\textrm{c}}, \Delta_{\textrm{set}},$ and $\tau/T_{\textrm{g}}$. The intuition for the first two parameters is that a large appliance population and allowable set point shift range enables the operator to provide more service. As for the parameter $\tau/T_{\textrm{g}}$, we note that a large value of $\tau$ impedes, and $T_{\textrm{g}}$ facilitates, the change of room temperature when the same control is applied. Thus the allowable accumulated provision is proportional to $\tau$ and inversely proportional to $T_{\textrm{g}}$.

When we consider short term regulation, the possible ramp of consumption is limited by the state $x(t)$ because we are shifting the set point rather than directly turning on/off appliances. Aside from $x(t)$, we are interested in finding a few fundamental parameters that characterize the short term capability similar to those found in the first proposition. The second proposition below provides similar results.

\textbf{Proposition.2} For a given number of appliances $N_{\textrm{c}}$ and width of temperature band for a specific set point $[T_{\textrm{set}}-\frac{1}{2}\Delta_{\textrm{band}},T_{\textrm{set}}+\frac{1}{2}\Delta_{\textrm{band}}]$, the amount of reserve that a building can provide for one period ramping is limited by the following bounds,
\begin{equation}
\label{them2}
-Nx_{\textrm{2N}}\frac{N_{\textrm{c}}T_{\textrm{g}}}{\tau\Delta_{\textrm{band}}}\leq \Delta R\leq Nx_{\textrm{N}}\frac{N_{\textrm{c}}T_{\textrm{g}}}{\tau\Delta_{\textrm{band}}},
\end{equation}
where $T_{\textrm{g}}$ is the temperature gain of air conditioner if it is on, and $\tau$ is the effective thermal constant of the building.

\textit{Proof.} In the dynamic operation when the tracking error is 0, the feedback controller is given by,
\begin{equation}
\label{uDeltaR}
u=\frac{1}{N_{\textrm{c}}(x_{\textrm{N}}+x_{\textrm{2N}})}[N_{\textrm{c}}(\alpha x_{\textrm{N}}-\beta x_{\textrm{2N}})-\Delta R].
\end{equation}
Since the allowable control set is $u\in [-\beta,\alpha]$, then $\Delta R$ is restricted by
\begin{equation}
\label{restriction2}
-N_{\textrm{c}}(\alpha+\beta)x_{\textrm{2N}}\leq \Delta R \leq N_{\textrm{c}}(\alpha+\beta)x_{\textrm{N}}
\end{equation}
Using the relation between transition rate and duty cycle developed in section \ref{markov jump process model} and the fact
\begin{displaymath}
\Delta_{\textrm{band}}=t_{\textrm{off}}r_{\textrm{off}}=t_{\textrm{on}}r_{\textrm{on}},
\end{displaymath}
the following inequality is seem to hold:
\begin{equation}
-N x_{\textrm{2N}}\frac{N_{\textrm{c}}(r_{\textrm{on}}+r_{\textrm{off}})}{\Delta_{\textrm{band}}} \leq \Delta R \leq N x_{\textrm{N}}\frac{N_{\textrm{c}}(r_{\textrm{on}}+r_{\textrm{off}})}{\Delta_{\textrm{band}}}.
\end{equation}
Using the fact that $r_{\textrm{on}}+r_{\textrm{off}}=T_{\textrm{g}}/\tau$ yields (\ref{them2}). $\blacksquare$

\textbf{Remark.2} According to proposition 2, the short term provision capability is proportional to $N_{\textrm{c}}$, and inversely proportional to $\Delta_{\textrm{band}}$ and  $\tau/T_{\textrm{g}}$. The proportionality to $N_{\textrm{c}}$ shares the same explanation as in the first remark. The two inverse proportionalities can be explained by saying that small value of $\Delta_{\textrm{band}}$ makes the width of individual temperature bin smaller, and large value of $T_{\textrm{g}}/\tau$ makes the thermal transfer faster. These two factors in turn facilitate the state transition to provide larger one-step service. 

\textbf{Remark.3} The advantage of taking comfort band $\Delta_{\textrm{band}}\rightarrow 0$ is that we can provide large value of one step service and at the same time make each room temperature stick to the set point, with the sacrifice that we shorten the appliance's duty cycle. This tradeoff between system performance and appliances functioning is consistent with what we find in \cite{bowen} where electricity consumption can be smoothed by shortening appliances duty cycle.

\textbf{Remark.4} The fraction $\tau/T_{\textrm{g}}$ affects both long and short term regulation provision. Based on (\ref{thm1}) and (\ref{them2}), it can be seen that a building able to provide large amount of short term ramping regulation is more incapable of providing long term accumulated regulation. The opposite also holds. Thus the two capabilities restrict each other.

\subsection{Buildings in the Regulation Service Market}
The above propositions determine the maximum regulation service that can be provided in the power market. To become a qualified provider in the U.S., a building needs to pass the \textbf{T-50} qualifying test \cite{PJM}. Fig.3 shows the test signal from PJM. The dotted line is the ideal consumption response that the building consumes according to the test signal. Two key requirements are needed to pass the test:

$\bullet$ Rate of response: the building needs to able to reach the maximum or minimum consumption level within 5 minutes.

$\bullet$ Sustained response: the building needs to able to maintain at the maximum or minimum consumption level for 5 minutes. 

\begin{figure}[htb]
\centering
\label{regulation_signal}
\includegraphics[width=0.4\textwidth,height=0.2\textheight]{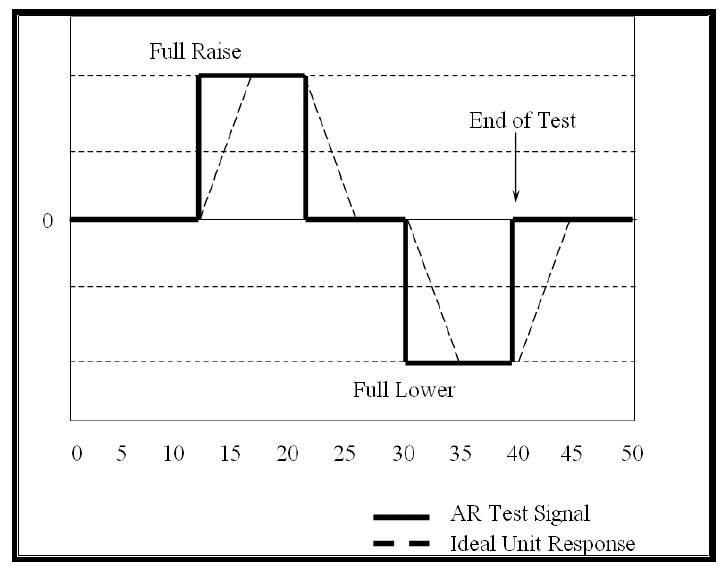}
\caption{T-50 test by PJM where regulation providers are obliged to ramp their consumption up and down to meet the test requirements.}
\end{figure}

The following corollary gives the upper bound on the regulation service that a building can provide.

\textbf{Corollary.} To pass the T-50 qualifying test with a rate of response within $k$ minutes $(k\leq 5)$, the maximum regulation service $R_{\textrm{r,max}}$ that a building can provide is given by,
\begin{equation}
\label{corollary bound}
R_{\textrm{r,max}}=\min \biggl\{\frac{N_{c}\tau\Delta_{\textrm{set}}}{20 T_{\textrm{g}}},\min(\frac{k N_{\textrm{c}}}{t_{\textrm{on}}}, \frac{k N_{\textrm{c}}}{t_{\textrm{off}}})\biggr\}.
\end{equation}

\textit{Proof.} For long term regulation service, the maximum value of $S(t)$ for $t\in[0,50]$ is $S(t)=10 R_{\textrm{r}}$ with $t=30$. According to proposition 1,
\begin{equation}
10 R_{\textrm{r}}\leq \frac{N_{\textrm{c}}\tau\Delta_{\textrm{set}}}{2T_{\textrm{g}}}.
\end{equation}
This yields
\begin{equation}
\label{long term}
R_{\textrm{r}}\leq \frac{N_{\textrm{c}}\tau\Delta_{\textrm{set}}}{20 T_{\textrm{g}}}.
\end{equation}
For short term regulation service, we need to track the rate of consuming up and down starting from steady state. It is easy to verify that the steady state probability distribution satisfies,
\begin{equation}
\left\{
\begin{array}{l}
x_{\textrm{1}}=\ldots=x_{\textrm{N}}=\frac{\displaystyle\beta}{\displaystyle N(\alpha+\beta)}\\
x_{\textrm{N+1}}=\ldots=x_{\textrm{2N}}=\frac{\displaystyle\alpha}{\displaystyle N(\alpha+\beta)}.
\end{array} \right.
\end{equation}
Substitute the above equation into (\ref{them2}) and note that $\Delta R= R_{\textrm{r}}/k$ in a $k$ minutes ramp process. Finally we have,
\begin{equation}
\label{short term}
R_{\textrm{r}}\leq \min(\frac{k N_{\textrm{c}}}{t_{\textrm{on}}},\frac{k N_{\textrm{c}}}{t_{\textrm{off}}}).
\end{equation}
Taking the minimum of (\ref{long term}) and (\ref{short term}) completes the proof. $\blacksquare$

\subsection{Real Time Spinning Reserve}
When the real time consumption $\Delta P$ ramps up or down quickly due to short term stochastic consumption aggregation, the ISO has to use spinning reserves to compensate for the demand that cannot be covered by the regulation service. This follow from the limits of proposition 1 and 2. In such cases, a feedback signal from the provider side to the ISO is beneficial to the ISO so that it can determine the needed spinning reserve $P_{\textrm{spin}}$. The following proposition establishes the relation between $P_{\textrm{spin}}$ and $\Delta P$ when necessary building information is given. 

\textbf{Proposition.3} Denote $\Delta P$ as the stochastic demand ramp in one step, then the spinning reserve $P_{\textrm{spin}}$ needed to maintain grid balance is given by,
\begin{equation}
\label{spinning}
\begin{array}{ll}
P_{\textrm{spin}}=&(\Delta P-\Delta R_{\textrm{max}})\mathbbm{1}_{\{\Delta P>\Delta R_{\textrm{max}}\}}+\\
&(\Delta P-\Delta R_{\textrm{min}})\mathbbm{1}_{\{\Delta P < \Delta R_{\textrm{min}}\}},
\end{array}
\end{equation}
where $\mathbbm{1}_{\{\cdot\}}$ is the indicator function, and
\begin{equation}
\label{Rvalues}
\setlength{\extrarowheight}{7pt}
\begin{array}{ll}
\Delta R_{\textrm{max}}=&N_{\textrm{c}}\biggl\{ (r_{\textrm{off}}x_{\textrm{N}}-r_{\textrm{on}}x_{\textrm{2N}})+\\
&\min(\frac{\displaystyle
T_{\textrm{set}}-T^{\textrm{min}}_{\textrm{set}}}{\displaystyle
\Delta t \delta},r_{\textrm{on}}) (x_{\textrm{N}}+x_{\textrm{2N}})\biggr\},\\
\Delta R_{\textrm{min}}=&N_{\textrm{c}}\biggl\{ (r_{\textrm{off}}x_{\textrm{N}}-r_{\textrm{on}}x_{\textrm{2N}})-\\
&\min(\frac{\displaystyle
T^{\textrm{max}}_{\textrm{set}}-T_{\textrm{set}}}{\displaystyle
\Delta t \delta},r_{\textrm{off}}) (x_{\textrm{N}}+x_{\textrm{2N}})\biggr\},
\end{array}
\end{equation}
are the maximum and minimum reserve provision threshold for the building.

\textit{Proof.} The allowable control $u$ is limited by two factors. The first is that the set point shift resulting from $u$ should be within the allowable set point range,
\begin{displaymath}
T_{\textrm{set}}+\Delta t \delta u \in [T_{\textrm{set}}^{\textrm{min}},T_{\textrm{set}}^{\textrm{max}}].
\end{displaymath}
The second is the requirement of maintaining non-negative Markov rate,
\begin{displaymath}
u\in[-\beta,\alpha].
\end{displaymath}
From the above two inequalities,
\begin{eqnarray}
\label{uallowablerange}
u\in\biggl[\max(\frac{T_{\textrm{set}}^{\textrm{min}}-T_{\textrm{set}}}{\Delta t\delta},-\beta),\min(\frac{T_{\textrm{set}}^{\textrm{max}}-T_{\textrm{set}}}{\Delta t\delta},\alpha)\biggr].
\end{eqnarray}
From (\ref{uDeltaR}), the one period ramping can be expressed in terms of $u$,
\begin{equation}
\label{deltaR}
\Delta R = N_{\textrm{c}}\big\{(\alpha x_{\textrm{N}}-\beta x_{\textrm{2N}}) - u(x_{\textrm{N}}+x_{\textrm{2N}})\big\}
\end{equation}
Substituting (\ref{uallowablerange}) into (\ref{deltaR}) yields $\Delta R \in [\Delta R_{\textrm{min}},\Delta R_{\textrm{max}}]$, where $\Delta R_{\textrm{min}}$ and  $\Delta R_{\textrm{max}}$ take values in (\ref{Rvalues}). For the grid balance, we have,
\begin{equation}
\Delta P=\Delta R+P_{\textrm{spin}},
\end{equation}
and we wish to use as little spinning reserve as possible. Then $P_{\textrm{spin}}$ will take value in (\ref{spinning}). The indicator function in (\ref{spinning}) gives the condition that $\Delta P$ is outside the range of $\Delta R$.$\blacksquare$

In the U.S. regulation service market, the ISO has purchased regulation service a day ahead from multiple $m$ providers. Assuming that feedback signals are set up between all the providers and the ISO such that the latter receives information from all of them every 4 seconds. Upon receiving these information, the ISO knows their individual instant capability range $[\Delta R_{\textrm{min}}^{i},\Delta R_{\textrm{max}}^{i}],\textrm{i=1,...,m}$, and then dispatches regulation signals to each of them. The signals are dispatched such that the minimum spinning generation is used and such that it should be within the range of capability for the building to respond. Most of the time, the one period ramping is within the total regulation capability of the $m$ buildings so no spinning generation is needed. In such cases, the ISO can have more than one dispatch solution. The question arises that how the ISO should dispatch signals in an \textit{optimized} way. The next subsection answers the question by solving a quadratic program $(QP)$.

\subsection{Optimization of Real Time Regulation Signal Dispatch}
When there is more than one solution to the regulation dispatch at time $t$ such that no spinning generation is needed, an optimal way to dispatch is that we maximize the regulation provision capability at time $t+1$. From proposition 3, the $i^{th}$ building can provide regulation service with range $\Delta R \in [\Delta R_{\textrm{min}}^{i},\Delta R_{\textrm{max}}^{i}]$. The width of the closed interval of regulation service is given by
\begin{equation}
\setlength{\extrarowheight}{7pt}
\begin{array}{ll}
W_{\textrm{d}}^{i}&=\Delta R_{\textrm{max}}^{i}-\Delta R_{\textrm{min}}^{i}\\
&=N_{\textrm{c}}^{i}(x_{\textrm{N}}^{i}(t+1)+x_{\textrm{2N}}^{i}(t+1))\\
&\biggl\{\min(\frac{\displaystyle
T_{\textrm{set}}^{i}(t+1)-T^{\textrm{min},i}_{\textrm{set}}}{\displaystyle
\Delta t \delta},r_{\textrm{on}}^{i})+\\
&\min(\frac{\displaystyle
T^{\textrm{max},i}_{\textrm{set}}(t+1)-T_{\textrm{set}}^{i}}{\displaystyle
\Delta t \delta},r_{\textrm{off}}^{i})\biggr\}.
\end{array}
\end{equation}
Then the objective function is given by
\begin{equation}
\textrm{max} \sum\limits_{i=1}^{m} W_{\textrm{d}}^{i} - M P_{\textrm{spin}}^{2},
\end{equation}
namely we are maximizing the sum of $m$ regulation service widths from each building at time $t+1$, minus the spinning generation penalty $P_{\textrm{spin}}$ with positive coefficient $M$. The maximization problem is subject to the following type of constrains:

\textbf{State Dynamics:}
\begin{equation}
\label{constraint:state dynamics}
\setlength{\extrarowheight}{7pt}
\begin{array}{l}
x_{\textrm{N}}^{i}(t+1)=x_{\textrm{N}}^{i}(t)+\Delta t (\alpha^{i}-u^{i})(x_{\textrm{N-1}}^{i}(t)-x_{\textrm{N}}^{i}(t)),\\
x_{\textrm{2N}}^{i}(t+1)=x_{\textrm{2N}}^{i}(t)+\Delta t (\beta^{i}+u^{i})(x_{\textrm{2N-1}}^{i}(t)-x_{\textrm{2N}}^{i}(t)).
\end{array}
\end{equation}

\textbf{Feedback Controller:}
\begin{equation}
N_{\textrm{c}}^{i}\Delta t (x_{\textrm{N}}^{i}(t)+x_{\textrm{2N}}^{i}(t)) u^{i}+\Delta R^{i}=N_{\textrm{c}}^{i}\Delta t (\alpha x_{\textrm{N}}^{i}(t)-\beta x_{\textrm{2N}}^{i}(t)).
\end{equation}

\textbf{Supply-demand Balance:}
\begin{equation}
\sum\limits_{i=1}^{m}\Delta R^{i} + P_{\textrm{spin}} = \Delta P.
\end{equation}

\textbf{Non-negative Markov Rate:}
\begin{equation}
\beta^{i} \leq u^{i} \leq \alpha^{i}.
\end{equation}

\textbf{Allowable Regulation Range:}
\begin{equation}
R_{\textrm{b}}-R_{\textrm{r}}\leq Cx + \Delta R^{i} \leq R_{\textrm{b}}+R_{\textrm{r}}.
\end{equation}

\textbf{Allowable Set Point Range:}
\begin{equation}
\label{constraint: set point allowable range}
T^{\textrm{min},i}_{\textrm{set}} \leq T_{\textrm{set}}^{i}(t+1)= T_{\textrm{set}}^{i}(t)+u^{i}\Delta t\delta \leq T^{\textrm{max},i}_{\textrm{set}}.
\end{equation}
The above optimization cannot be solved with standard technique because the objective function has the minimum operator. We transform the original problem into a $QP$. Let
\begin{equation}
\label{45}
\setlength{\extrarowheight}{7pt}
\begin{array}{l}
m_{1}^{i}=\min(\frac{\displaystyle
T_{\textrm{set}}^{i}(t+1)-T^{\textrm{min},i}_{\textrm{set}}}{\displaystyle\Delta t \delta},r_{\textrm{on}}^{i}),\\
m_{2}^{i}=\min(\displaystyle\frac{T^{\textrm{max},i}_{\textrm{set}}-T_{\textrm{set}}^{i}(t+1)}{\displaystyle\Delta t \delta},r_{\textrm{off}}^{i}).
\end{array}
\end{equation}
Then the original objective function becomes
\begin{equation}
\label{new qp}
\setlength{\extrarowheight}{7pt}
\begin{array}{ll}
\textrm{max} & \sum\limits_{i=1}^{m} N_{\textrm{c}}^{i}(x_{\textrm{N}}^{i}(t+1)+x_{\textrm{2N}}^{i}(t+1))(m_{1}^{i}+m_{2}^{i})\\
& - M P_{\textrm{spin}}^{2}.
\end{array}
\end{equation}
If we add the following constrains,
\begin{equation}
\setlength{\extrarowheight}{7pt}
\label{constraint: additional m}
\begin{array}{l}
m_{1}^{i}\leq r_{\textrm{on}}, \\
m_{1}^{i}\leq \frac{\displaystyle T_{\textrm{set}}^{i}(t+1)-T^{\textrm{min},i}_{\textrm{set}}}{\displaystyle\Delta t \delta},\\
m_{2}^{i}\leq r_{\textrm{off}}, \\
m_{2}^{i}\leq \frac{\displaystyle T^{\textrm{max},i}_{\textrm{set}}-T_{\textrm{set}}^{i}(t+1)}{\displaystyle\Delta t \delta},
\end{array}
\end{equation}
and solve the $QP$,
\begin{equation}
\setlength{\extrarowheight}{7pt}
\label{final QP}
\begin{array}{ll}
\max &(\ref{new qp})\\
\textrm{s.t.} & (\ref{constraint:state dynamics})-(\ref{constraint: set point allowable range}), (\ref{constraint: additional m}),
\end{array}
\end{equation}
we will obtain the same optimal solution as solving the original problem. This is because when reaching the optimal solution, one of the inequality constraints for both $m_{1}^{i}$ and $m_{2}^{i}$ in (\ref{constraint: additional m}) will be strict, otherwise the optimal solution is not reached since we can increase the value of $m_{1}^{i}$ or $m_{2}^{i}$ to increase the value of objective function due to positive coefficient $N_{\textrm{c}}^{i}(x_{\textrm{N}}^{i}(t+1)+x_{\textrm{2N}}^{i}(t+1))$ in (\ref{new qp}). Then (\ref{45}) is satisfied and optimizing over (\ref{final QP}) is equivalent to solving the original problem. Note that (\ref{final QP}) has quadratic objective function with linear constraints, Matlab or CPLEX can solve this $QP$ efficiently.

\section{State Observer Design}
\label{state observer design}
In controller design (\ref{controlaw}), it is assumed that $x(t)$ can be measured. If this is not true, especially when the temperature band is finely discretized, i.e. $\delta$ is small so that the sensor cannot provide the required precision, we need to design an observer to estimate the state. Considering the following dynamics of the observer,
\begin{equation}
\dot{\tilde{x}}=A\tilde{x}+B\tilde{x}u+L(t)(y-C\tilde{x}),
\end{equation}
where $L(t)$ is the time varying column vector to be designed. The last term is similar to the innovation term in Luenberger filter \cite{kalmanfilter}. Define the estimation error, $e=x-\tilde{x}$, then
\begin{equation}
\dot{e}=(A+Bu-LC)e.
\end{equation}
In the observer design, we restrict the allowable control set to prevent control saturation with some small constant positive margin $\tilde{\epsilon}>0$,
\begin{equation}
\label{allowable control set}
u(t)\in[-\beta+\tilde{\epsilon}\leq u(t)\leq \alpha-\tilde{\epsilon}].
\end{equation}
This guarantees an irreducible and ergodic Markov chain. If the control signals is allowed to stay saturated, then the Markov chain breaks and some of the states become isolated, see Fig.1(a) when you substitute $u=-\beta$ or $u=\alpha$ into the model. Estimation error of those isolated states cannot be guaranteed. 

The following proposition states that we can select $L(t)$ such that the estimation error will approach zero as $t\rightarrow\infty$.

\textbf{Proposition.4} If the control is chosen according to (\ref{allowable control set}), then the following holds:

(1) There exists a time varying observer design $L(t)=[\underbrace{0,\ldots,0}_{2N-1},L_{\textrm{2N}}(t)]^{T}$ such that 
\begin{equation}
\label{them4_1}
e^{T}(A+Bu-LC)e < -\epsilon(t) \|e\|^{2}
\end{equation}
for all $e(t)$ with $\epsilon(t)=\gamma\min [\beta+u(t),\alpha-u(t)]$, $\gamma\in (0,1)$. 

(2) The estimation error would converge to zero asymptotically 
\begin{equation}
\lim\limits_{t\rightarrow\infty}e(t)=0
\end{equation}

\textit{Proof.} (1). To prove (\ref{them4_1}), it is equivalent to show that
\begin{equation}
e^{T}(A+Bu-LC+\epsilon(t)I)e < 0.
\end{equation}
Denote $\tilde{A}=A+Bu-LC+\epsilon(t)I$, then $\tilde{A}(t)$ can be expressed by the following equation based on the special structure of the state space matrices $A,B$ and $C$ derived in section \ref{markov jump process model}.

{\scriptsize
\begin{equation}
\label{tiledeA}
\begin{array}{ll}
\tilde{A}=\\
\left( \begin{array} {ccccccc}
-\alpha+u & 0 & &\ldots &  & & \beta+u \\
\alpha-u & \ddots & & & & & \\
0 & \ddots & -\alpha+u &\ddots &  & & \\
& & \alpha-u & -\beta-u &  & &\vdots \\
\vdots& & \ddots& \beta+u & \ddots & & \\
& & & &\ddots & -\beta-u & 0 \\
0 & & \ldots & 0  & \ldots& \beta+u &-\beta-u
\end{array} \right)+\\
\left( \begin{array} {ccccccc}
\epsilon(t) & 0 & &\ldots &  & & 0 \\
0 & \ddots & & & & & \\
0 & \ddots & \epsilon(t) &\ddots &  & & \\
& & 0 & \epsilon(t) &  & &\vdots \\
\vdots& & \ddots& 0 & \ddots & & \\
& & & &\ddots & \epsilon(t) & 0 \\
0 & \ldots & 0 & -N_{c}L_{2N}  & \ldots& -N_{c}L_{2N} &\epsilon(t)-N_{c}L_{2N}.
\end{array} \right).
\end{array}
\end{equation}
}

Denote $\tilde{A}_{i}$ as the $i-$by$-i$ square matrix from the upper left part of $\tilde{A}$. In order to show that $\tilde{A}$ is negative definite, it is equivalent to show that 
\begin{equation}
(-1)^{i}\det(\tilde{A}_{i}) > 0, \textrm{for $i=1,\ldots, 2N$}.
\end{equation}
For $i=1,\ldots,2N-1$, $\tilde{A}_{i}$ is a triangular matrix whose determinant is the product of its diagonal elements. From (\ref{tiledeA}), the diagonal elements $\tilde{A}_{i,i}$ is given by,
\begin{equation}
\tilde{A}_{i,i}= \left\{
\begin{array}{ll}
-\alpha+u+\epsilon(t) & \textrm{if $i=1,\ldots,N$} \\
-\beta-u+\epsilon(t) & \textrm{if $i=N+1,\ldots,2N-1$}
\end{array}
\right.
\end{equation}
Since $\epsilon(t)=\gamma\min [\beta+u(t),\alpha-u(t)]$, we have $A_{i,i}\leq (\gamma-1)\min [\beta+u(t),\alpha-u(t)] < 0$ for $i=1,\ldots,2N-1$. This yields
\begin{equation}
(-1)^{i}\det(\tilde{A}_{i})=(-1)^{i}\prod\limits_{j=1}^{i}\tilde{A}_{j,j}> 0, \textrm{for $i=1,\ldots,2N-1$}.
\end{equation}
As for $i=2N$, $\det(\tilde{A})$ is a linear function of $L_{2N}$ as the observer parameter only appears in the last row of the matrix, so there must exists a $L_{2N}$ such that $\det(\tilde{A})> 0$. This completes the proof of negative definiteness of the matrix $\tilde{A}$, and (\ref{them4_1}) holds.

(2). Let 
\begin{equation}
\label{lyapunov y}
y(t)=e^{\int\limits_{0}^{t}\epsilon(\tau)d \tau}\|e(t)\|^{2},
\end{equation}
then the time derivative of $y(t)$ is given by,
\begin{equation}
\setlength{\extrarowheight}{7pt}
\label{y upper bound}
\begin{array}{ll}
\dot{y}(t)&=\epsilon(t)e^{\int\limits_{0}^{t}\epsilon(\tau)d \tau}\|e(t)\|^{2}+2e^{T}(t)\tilde{A}e(t)e^{\int\limits_{0}^{t}\epsilon(\tau)d \tau}\\
&=e^{\int\limits_{0}^{t}\epsilon(\tau)d \tau}[\epsilon(t)\|e(t)\|^{2}+2e^{T}(t)\tilde{A}e(t)]\\
&< e^{\int\limits_{0}^{t}\epsilon(\tau)d \tau}[\epsilon(t)\|e(t)\|^{2}-2\epsilon(t)\|e(t)\|^{2}]\\
& = -e^{\int\limits_{0}^{t}\epsilon(\tau)d \tau} \epsilon(t)\|e(t)\|^{2} \\
& =-\epsilon(t)y(t),
\end{array}
\end{equation}
The time derivative from the above equation indicates that $y(t)$ in (\ref{lyapunov y}) is stable at zero. 

On the other hand, from (\ref{lyapunov y}) we have $\lim\limits_{t\rightarrow\infty}e^{\int\limits_{0}^{t}\epsilon(\tau)d \tau}=\infty$ and $\lim\limits_{t\rightarrow\infty}y(t)=0$. Necessarily we need $\lim\limits_{t\rightarrow \infty}e(t)=0$ to satisfy these two conditions, which indicates that the error will approach zero asymptotically.$\blacksquare$

\textbf{Remark.5} $\epsilon(t)=\gamma\min [\beta+u(t),\alpha-u(t)]\geq \gamma\tilde{\epsilon}$ is the lower (conservative) bound of the convergence rate we can achieve. In real time operation when the control signal is distributed across the allowable set to be away from saturation, the numerical convergence rate is larger than the conservative bound. It should be noted that as the duty cycle increases, the boundary value of $\alpha=\frac{\displaystyle N}{\displaystyle t_{\textrm{off}}}$ and $\beta=\frac{\displaystyle N}{\displaystyle t_{\textrm{on}}}$ decreases, which means we have smaller allowable control set. Then the value of control signal would be closer to the boundary than the scenario with small duty cycle appliances, therefore it is anticipated that the numerical convergence speed of the error vector will be slower for system having large duty cycle appliances.
\section{Simulation}
In this section, we provide three simulation to verify the theoretical framework proposed in previous sections. In the first example, we will show how building parameters derived in the first two propositions restrict the maximum delivery of regulation service. In the second one, we will verify the effectiveness of the proposed $QP$ optimization in regulation signal dispatch by reducing real time spinning generation. In the last example, we will show the performance of the observer that asymptotically reduces the estimation error.
\label{simulation section}
\subsection{Long and Short Term Regulation Service Limitation}
Fig.4 shows how the long and short term restrictions on regulation service affect the T-50 test. The first figure is an example of the inability to provide enough accumulated regulation service due to limited allowable set point choice. In this figure, the set point shift hits the lower bound of the allowable set between 15 to 20 minutes. As a consequence, the building cannot provide a sustained high consumption level by further decreasing the set point. Similarly, the set point shift hit the upper bound after 40 minutes which restricts the sustain provision of lower level aggregated consumption.  

The second figure is an example to show the inability to provide short term regulation service. Although the set point shift is within the allowable set, the ramping speed to provide positive and negative regulation service is smaller than the required speed for a given maximum regulation obligation. This is because the thermostatic appliances in this simulation have large duty cycle. For example, the typical duty cycle of refrigerators is around 20 minutes and has large effective thermal constant \cite{Cavallo}. Such a large duty cycle greatly restricts the maximum ramping rate in regulation service derived from proposition 2.
\begin{figure}[hbt]
\label{T-50 test}
\centering
\subfigure[Long Term Provision]{
\includegraphics[width=0.4\textwidth,height=0.2\textheight]{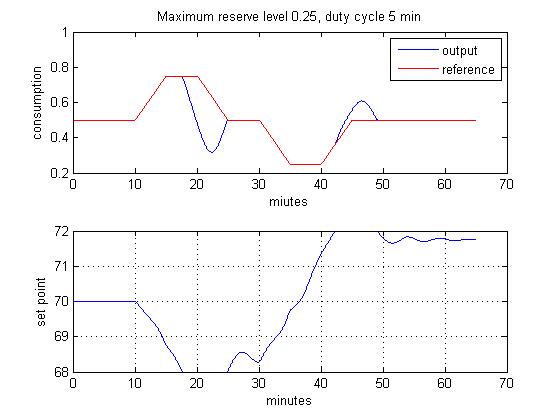}
}
\subfigure[Short Term Provision]{
\includegraphics[width=0.4\textwidth,height=0.2\textheight]{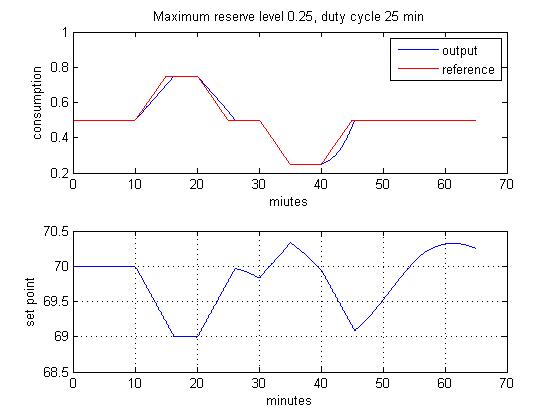}
}
\caption{Individual providers fail to pass the T-50 qualifying test due to either a bounded allowable set point range or a limited ramping capability.}
\end{figure}
\subsection{Optimal Regulation Signal Dispatch}
We use real time PJM data \cite{realsignal} to verify the performance of the overall two level control and optimization framework. We can see that the regulation signal is stochastic and ramps up and down according to the area control error with no apparent probability distribution. Fig.5 shows the simulation results of the proposed system with lower level feedback linearization controller design and higher level two-way communication feedback design. Fig.6 is the simulation result with only lower level controller. We find that both systems can track the regulation signal when it ramps up or down with the design of building level feedback controller, but they differ in the maximum ability of tracking. Compared with Fig.6, where individual building regulation signals are proportional to their maximum provision level (ISO does not have knowledge about real time building information, so it dispatches signals proportional to their a prior maximum commitment), the signals in Fig.5 are determined by solving the $QP$. We find that the consumption trend of red curve and blue curve are different most of the time in Fig.5 that fully exploits their tracking capability, while in Fig.6 individual building consumptions follow the same trend with proportionality. The accumulated generation of spinning reserves is reduced when the regulation dispatch is optimized by the $QP$, and the distribution of spinning reserve is more concentrated around $0$ with smaller variance. Tab.1 shows the statistics of both systems, the total spinning generation is reduced by $50\%$ after optimization. The standard deviation, maximum, and minimum spinning generation is reduced by $30\%$, $30\%$, and $15\%$ respectively. Furthermore, at times when the regulation signal ramps up and down at a fast speed, for example at 15, 30, 35, 40, and 50 minutes, the performance of the two level structured system in Fig.5 is better than the system performance without the design. This is because the required spinning reserves is largely reduced despite the sharp ramping regulation signal. To sum, the proposed two level feedback framework outperforms the system with only building level controller design.
\begin{figure}[hbt]
\centering
\subfigure[Building Outputs]{
\includegraphics[width=0.4\textwidth,height=0.2\textheight]{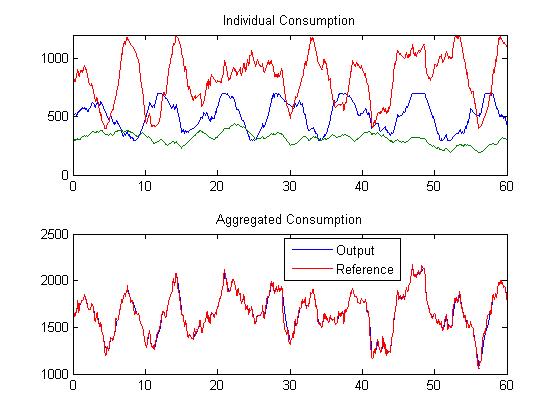}
}
\subfigure[Spinning Generation]{
\includegraphics[width=0.4\textwidth,height=0.2\textheight]{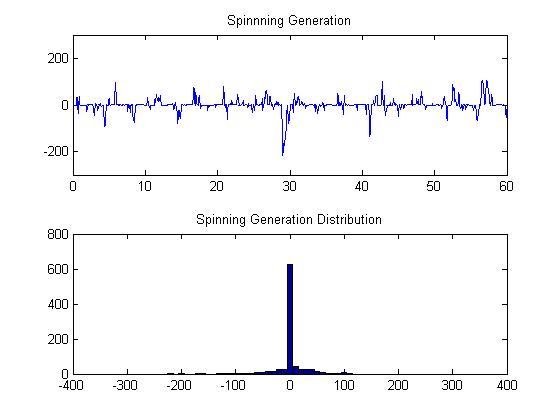}
}
\label{opt_result}
\caption{Real time regulation dispatch after the quadratic program is solved. Tracking becomes close to the target and spinning generation is more concentrated around 0 with smaller variance.}
\end{figure}
\begin{figure}[hbt]
\centering
\subfigure[Building Outputs]{
\includegraphics[width=0.4\textwidth,height=0.2\textheight]{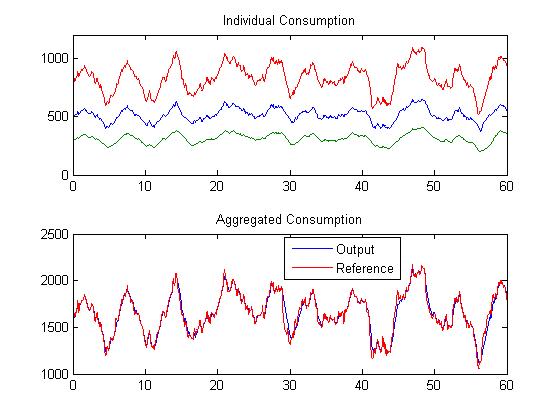}
}
\subfigure[Spinning Generation]{
\includegraphics[width=0.4\textwidth,height=0.2\textheight]{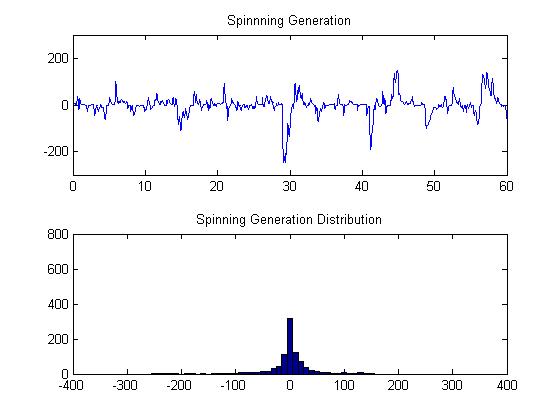}
}
\label{propor_result}
\caption{Real time regulation dispatch when the ISO does not receive information from individual providers. Spinning generation is more spread out around 0 with larger variance.}
\end{figure}
\begin{table}[hbt]
\caption{Comparison of System Statistics}
\centering
\begin{tabular}{|c|c|c|c|c|c|}
\hline Unit/kW & Spin. Gen. & Mean & Std. Dev. & Max & Min \\ 
\hline Opt. Alg. & 10216 & 0.18 & 27.20 & 100.27 & -210.59 \\ 
\hline Prop. Alg. & 20525 & -0.44 & 41.97 & 146.19 & -247.65 \\
\hline
\end{tabular} 
\end{table}

\subsection{Observer Performance}
We verify the performance of the proposed observer. Fig.7(a) shows the real and estimated value of two sample states, $x_{\textrm{N}}$ and $x_{\textrm{2N}}$, these are the two states around the boundary of the comfort band that will affect the change in electricity consumption. We can see that the estimation error for these two boundary states approach zero after 20 minutes. Fig.7(b) shows the convergence speed of the norm decay of the error vector $e(t)$ for appliances with different duty cycles. The norm converges to zero after 20 minutes when the duty cycle is 10 minutes, and converges to zero after 35 minutes when the duty cycle is 20 minutes. This is because smaller duty cycle enables a larger allowable control set, i.e the system has large values of $\alpha$ and $\beta$ to prevent from saturation. In Fig.7(c) and Fig.7(d) we compare the distribution of control signals across the allowable control set, it can be seen that the control signal stays further away from saturation for system having smaller duty cycle. Such systems have a larger convergence speed of the error vector. 
\begin{figure}[hbt]
\centering
\subfigure[State Estimation]{
\includegraphics[width=0.4\textwidth,height=0.2\textheight]{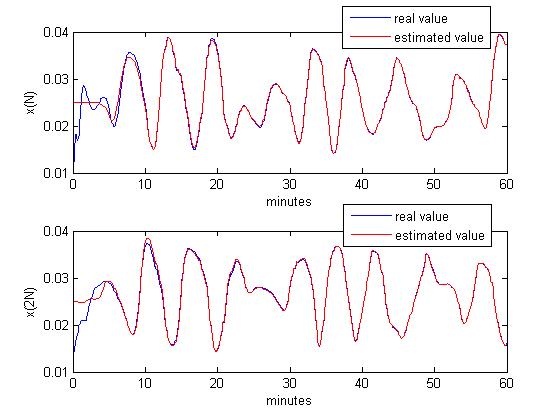}
}
\subfigure[Norm of the Error Vector]{
\includegraphics[width=0.4\textwidth,height=0.2\textheight]{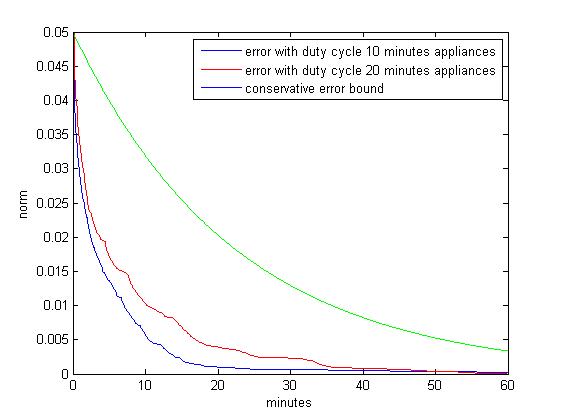}
}
\subfigure[Control Distribution with Duty Cycle 10 Minutes]{
\includegraphics[width=0.4\textwidth,height=0.2\textheight]{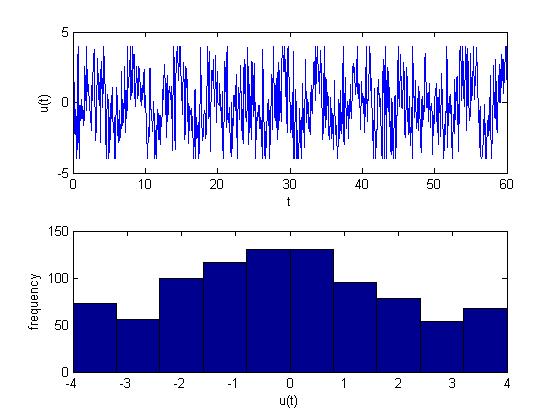}
}
\subfigure[Control Distribution with Duty Cycle 20]{
\includegraphics[width=0.4\textwidth,height=0.2\textheight]{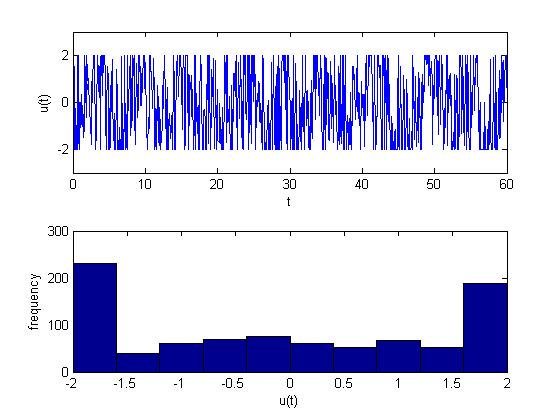}
}
\label{observer_result}
\caption{Observer performance validation. (a) Real and estimated value of two sample states $x_{\textrm{N}}$ and $x_{\textrm{2N}}$. (b).The norm of the estimation error gradually approaches zero as the estimation proceeds. Convergence rate of the estimation error vector depends on the duty cycle value. (c) and (d) the distribution of control signal across the allowable set when different duty cycle values are used}
\end{figure}

\section{Conclusions and Future Work}
\label{conclusion section}
This paper proposes an innovative two level feedback design system for the real time regulation provision. The lower level building feedback controller allows the building operator to track a given signal from the ISO, and the higher level information feedback from individual buildings to the ISO allows the latter to optimally dispatch real time regulation signals to multiple providers by solving a quadratic programming problem such that the cost of spinning reserves is reduced. During the development of the system, we also derive analytically the building's limitation on both long term and short term regulation provision and provide intuitive explanations that match our theoretical results. To counter the problem with effective measuring, we also proposed an observer design to reduce the error of state estimation such that the feedback controller can work well. Simulation results indicate that the proposed framework is superior in reducing the amount of real time spinning reserve.

Future work will include the modelling of similar two level feedback system but with price based signalling between the operator and the consumers. Such design will fit into problems where consumers do not authorize the operator to directly control their set point. It is expected that price-based mechanism has looser control compared with the direct load control because consumers can choose their own utility function. Consequently, it enables consumers to obtain different level of desired comfort.

\end{document}